# PERFORMANCE CONSTRAINT AND POWER-AWARE ALLOCATION FOR USER REQUESTS IN VIRTUAL COMPUTING LAB

**Nguyen Quang Hung, Nam Thoai, Nguyen Thanh Son**

*Ho Chi Minh City University of Technology, Vietnam*

Corresponding author: *hungnq2@cse.hcmut.edu.vn*

## ABSTRACT

Cloud computing is driven by economies of scale. A cloud system uses virtualization technology to provide cloud resources (e.g. CPU, memory) to users in form of virtual machines. Virtual machine (VM), which is a sandbox for user application, fits well in the education environment to provide computational resources for teaching and research needs. In resource management, they want to reduce costs in operations by reducing expensive cost of electronic bill of large-scale data center system. A lease-based model is suitable for our Virtual Computing Lab, in which users ask resources on a lease of virtual machines. This paper proposes two host selection policies, named MAP (minimum of active physical hosts) and MAP-H2L, and four algorithms solving the lease scheduling problem. FF-MAP, FF-MAP-H2L algorithms meet a trade-off between the energy consumption and Quality of Service (e.g. performance). The simulation on 7-day workload, which converted from LLNL Atlas log, showed the FF-MAP and FF-MAP-H2L algorithms reducing 7.24% and 7.42% energy consumption than existing greedy mapping algorithm in the leasing scheduler Haizea. In addition, we introduce a ratio θ of consolidation in HalfPI-FF-MAP and PI-FF-MAP algorithms, in which θ is $\pi/2$ and $\pi$, and results on their simulations show that energy consumption decreased by 34.87% and 63.12% respectively.

# 1. INTRODUCTION

Energy-efficient resource management is considered in parallel and distributed computing domain in recent years as Green computing [2], [3]. Cloud computing, which is popular with "pay-as-you-go" utility model [4], is economy driven. Saving operating costs (e.g. electric bill) of cloud system is necessary in any cloud provider. Energy used in servers and cooling in data centers of cloud system is costly in electric bill. In report of United States Environmental Protection Agency (EPA), it said that energy consumption of only federated servers and data centers in this nation will be reached 100 billion KWh in 2011 [5]. Therefore, reducing energy consumption in servers is important.

Energy-efficient resource management in cloud system is a hot topic. The problem is trade-off between minimizing of energy consumption and satisfying QoS (e.g. performance, resource availability on time for reservation request, etc.). Resource requirements depend on the applications and we are interested in the virtual computing lab, which is a cloud system to provide resources for teaching and researching. We consider three following cases using computational resources:

i. For teaching, a teacher in a college/university needs a cluster of tens of computing nodes to teach a MPI class for thirty students; it is a repeated event on Thursday, from 9:00 AM to 11:00 AM for about ten weeks. This is a kind of advanced reservation (AR) requests.

ii. For researching, a researcher wants executing a program with 10 nodes, each with 2 CPUs and 2GB of memory, with or without deadline on the job. This is a kind of best-effort (BE) requests.

iii. For self-study, a student wants virtual machines to do exercises. This is a kind of weak advanced reservation request which available time can shift on weekdays.

In this paper, we present four power-aware allocation algorithms for user lease request into computational resources, which are networked computers (physical hosts). The work [1] has argued that such scheduling is NP-Complete. To solve the assignment problem, this paper presents two host selection policies (MAP and MAP-H2L) and First-Fit mapping heuristics in using least physical hosts on mapping and shutting down free workload hosts to save energy. These four power-aware allocation algorithms (FF-MAP, FF-MAP-H2L, HalfPI-FF-MAP, PI-FF-MAP) are based on lease-based model presented in [1]. No scheduling algorithm addressed in power-aware scheduling/allocation supports lease-based model. Our power-aware allocation algorithms have implemented in scheduler of a resource lease manager (e.g. Haizea [6]). Our work did not use voltage and frequency scaling (e.g. Dynamic Voltage and Frequency Scaling) as in [7]. In addition, our work differs from [8], [9] in that our proposed architecture accepts lease-based model as in [1] for provision resources to user on both the best-effort and advanced reservation lease requests in teaching and researching cases, and our mapping algorithms avoid migrating of virtual machines.

This paper also proposes a ratio $\theta$ of consolidation, to map up to $\theta$ times more resource requests than capacity of resources of host. Purpose of this mechanism is to decrease energy consumption. On our simulation, in case the ratio $\theta$ is 1, $\pi/2$ and $\pi$ then a simulated data center with 1000 physical servers serves 10 days decreases in 116.63 KWh (7.24%), 561.51 KWh (34.87%), and 1016.61 (63.12%) respectively. The effect of this consolidation, however, is service level agreement (SLA) with customers can be violated.

The paper is structured as follows: In Section 2, we will provide a background on virtualization and simulation tool (e.g. Haizea). Related works are presented in Section 3 and the system architecture is introduced in Section 4. Section 5 is energy model. Our four power-aware allocation algorithms are presented in Section 6. Then in Section 7, we discuss on experimental study on the four algorithms. The last section is conclusion and future works.

## 2. BACKGROUND

### 2.1. Virtual Machine

Virtual machine [11], [12] is an entity, which is executed and managed by a Virtual Machine Monitor (VMM) or hypervisor such as Oracle VirtualBox, VMware, XEN, KVM, etc.

### 2.2. Haizea

Haizea [13] software, which is a resource lease manager, implements for the above lease-based model [1]. The existing scheduling algorithms in Haizea are greedy and map as much as possible virtual machines in a node up to maximum of its capacity. The scheduling algorithm just concerns on waiting time, preemption/non-preemption for best effort and advanced reservation leases. Our policy is new plug-in for Haizea to toward the using as less used servers as possible.

## 3. RELATED WORKS

B. Sotomayor's PhD thesis [1] presents the lease-based model which uses leases and virtual machines abstraction for user need requests. His thesis [1] contributes Haizea software [6]. Role of Haizea as a lease resource manager and virtual machine scheduler for OpenNebula, which is a cloud management software, is discussed in [14]. On works [1], [15-17], he develops greedy mapping algorithm for a lease $l$ at a time $t$ and duration in $d$ seconds, the greedy mapping algorithm sorts the physical hosts in order that lowest leases scheduled on them at time $t$ firstly and a free workload physical host prefers first chosen one in the list of physical hosts. Our MAP host selection policy sorts these free workload hosts in the tail of the list of physical hosts instead, and our mapping algorithm for virtual machines (VMs) is First-Fit heuristic. In other power-aware algorithms in [9], [18], [19], they showed that using less physical hosts can save energy consumption. In this paper, we address on heuristics in power-aware allocation of VMs by minimizing the number of working hosts. The paper presents the MAP host selection policy and First-Fit mapping heuristic to do such that. The MAP, MAP-H2L and First-Fit mapping heuristic have installed as plug-in in Haizea software for evaluation. Our power-aware allocation algorithms on simulation showed that they all reduced energy consumption.

There are some approaches on power-aware allocation/scheduling algorithm. In [7], they use technique of Dynamic Voltage Frequency Scaling (DVFS) to save energy consumption on CPU in cluster on scheduling. In the light workload, the system puts low frequency in CPU to reduce energy. Another approach [8], [9] is using Best-Fit heuristic on placement of virtual machines on data-center is presented. These authors used CloudSim software to evaluate of their allocation algorithm. Our power-aware allocation algorithms differ from placement algorithms in [8], [9] in: (i) we sort list of VMs order by duration and list of active hosts order by ranking score which depends on number of current leases on each host; (ii) assignment ends with all requested resources on each lease is satisfied successfully. In addition, these previous works [7-9] concern only on batch jobs or only immediate allocation of virtual machines. Even they did not concern on arrival time of requests. In the other hand, they did not address to both best-effort (BE) and advanced reservation (AR) leases on the same system.

Another mathematic programming approach to allocation of virtual machines is concerned in [10]; authors give Integer Liner Programming (ILP) equalization. To solve the ILP model, they use some strategies such as First-fit and Best-fit heuristics on the Bin Packing Problem. However, the ILP model does not address to arrival time of user requests, or advanced reservation lease, neither they have objective to performance constraint such as waiting time or slowdown.

## 4. SYSTEM ARCHITECTURE

We propose system architecture for an energy-efficient resource manager of private cloud. The system uses software such as OpenNebula and Haizea to set up a private cloud to provision resources for user requests. Fig. 2 shows the proposed system architecture (Fig. 2a) and VM scheduler (Fig. 2b) for provision resources. We discuss in detail about the VM scheduler with power-aware allocation in following sections.

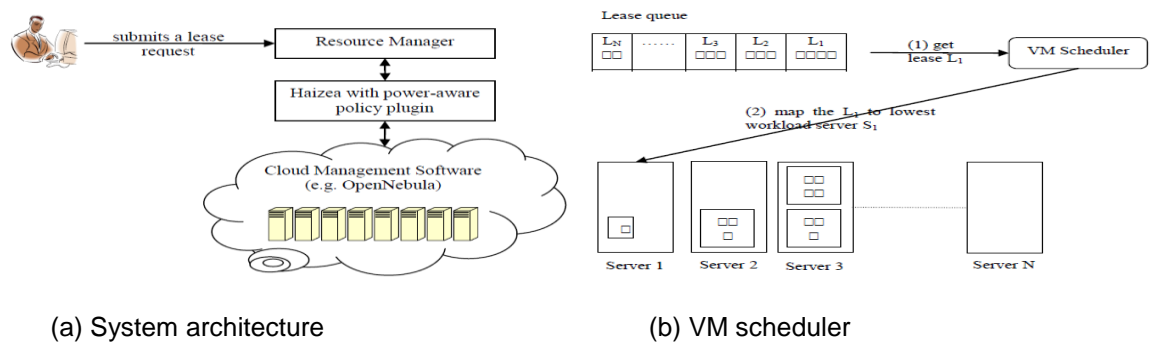

(a) System architecture  (b) VM scheduler

*Figure 1.* Proposed system architecture (a) and VM scheduler (b) for provision resources.

## 5. ENERGY MODEL

Xiaobo Fan et al. presented their work on power estimation that implies that peak power of CPU component is linear (e.g., one CPU peak power is 40 Watt, two CPU peak power is 80 Watt) [20]. They assume that CPU utilization as main factor in their power model and present an approximation for total system power against CPU utilization (u). A relaxed linear model is:

$$\text{Total system power} \approx P_{idle} + (P_{busy} - P_{idle}) \cdot u \quad (1)$$

With full (100%) CPU utilization ($u=1$), the power of the system $P_{busy}$ is called $P_{max}$ (Watt), give $k$ is ratio on power of the idle system against full utilization system:

$$k = P_{idle} / P_{max} \quad (2)$$

We assume that power is calculated as in [8] with k=0.7, in which a physical host is just 30 percents in different in power of in free CPU utilization and in full (100%) CPU utilization. Therefore, power of a physical host (P) is approximately by CPU utilization ($u$) as:

$$P \approx k \cdot P_{max} + (1-k) \cdot P_{max} \cdot u \approx P_{max}(0.7 + 0.3 \cdot u) \quad (3)$$

The power is calculated by time $t$ of the physical host is:

$$P(t) \approx P_{max}(0.7 + 0.3 \cdot u(t)) \quad (4)$$

In which $0 \leq u(t) \leq 1$: u(t) is ratio of number of working CPU cores and total CPU cores on a physical host. We assume that u(t) is unchanged between two events $t \in [t_{e1}, t_{e2}]$ of VM scheduler. Then the energy consumption calculates as in [8].

We assume that scheduler providing CPU utilization for user lease is steady on duration time.

## 6. POWER-AWARE SCHEDULING ALGORITHM

A cloud system is a set of homogeneous physical hosts, *pnodes*={$h_0, h_1,..., h_n$} with n≥0. Each physical machine includes many cores and so it is able to contain many virtual computers at the same time. There are two modes for any physical machine: (1) *active mode* when there is at least one running virtual machine on it, and (2) *passive mode* when there is no running virtual machine on it and so we will turn off these computers to reduce the energy. A computer in active/passive mode is sometimes called active/passive host.

Users request on leases, list of leases, *leases*={$l_1, l_2,..., l_m$}, each lease *l* has set of leasable nodes, *vnodes*[*l*] (e.g. each node is a virtual machine). Information about each lease includes arrival time, start time, duration in seconds, value of each needed resource type (e.g. CPU, memory, network bandwidth), virtual machine image disk file (including two required attributes are id and size), whether considers transferring image or not.

Our trade-off algorithm of performance and power-aware allocation of virtual machines tries to map as much as possible leases onto *active* hosts (i.e. the physical hosts have any assigned lease) and concern on minimize the number of active physical hosts. This algorithm assumes that an advanced reservation lease preempts all other best-effort leases.

Our customized host selection policy, named MAP, differs from existing host selection policy of Haizea in ranking formula. Our ranking formula prefers to active hosts to assign the new coming leases. Please note that idle (passive) host is always in the end of ordered list (the host's score is negative number).

The MAP-H2L host selection policy is similar to the MAP, but in contrast to the MAP, the MAP-H2L sorts list of physical hosts by highest workload to lowest workload.

Our First-Fit mapping differs the original greedy mapping algorithm in two points: (1) we sort list of physical hosts by using the above MAP (or MAP-H2L) host selection policy; (2) we choose the first matching active host $h \in \{h_0, h_1, ..., h_k\}$ ($\forall k: 0 \leq k \leq n$) that can fit at least full requested resources of one node in node set (nodes[$l$] VMs) of the lease $l$. We assigned up to maximum of the physical host $h$.

## 7. EXPERIMENTAL STUDY ON POWER-AWARE ALLOCATION

### 7.1. Workloads

We used a the LLNL Atlas log [21] from a Linux cluster at LLNL, and run a script, which is provided from the technical contribution of the thesis by Borja Sotomayor (2010) [1] to extract and convert requests in the log file to lease request tracefile to simulate on Haizea software. During the converting, information on the number of jobs, the job arrival time, time to finish the job (to calculate the time duration that the job is executed) are unchanged.

We extracted requests of ten days on LLNL Atlas log file and creates Haizea's lease tracefile (LLNL-s10d-small.lwf file), in which it has total of 1750 leases (all of them are scheduled successfully).

### 7.2. Simulated cluster

The simulated cloud data center has 1000 homogeneous, physical hosts. Each host has 8-cores CPU. The time of arrival of requests, the time duration of each lease and even the number of virtual machines to each lease are different in general.

### 7.3. Experiments

The simulation clock started at 2011-06-02 13:36:45.00 and stopped at 2011-05-23 07:00:00.00. We concern on CPU utilization to calculate their energy consumption of only the active physical hosts, we assume that other passive physical hosts do not consume energy. After running the simulation by Haizea on data from the LLNL-s10d-small.lwf file. We experimented with four algorithms allocation for VMs:

(1) **OriginalGreedy** algorithm: The original Greedy algorithm in Haizea,

(2) **FF-MAP** algorithm: The First-Fit heuristic in the VM scheduler and MAP host selection policy to minimize the number of active physical hosts,

(3) **FF-MAP-H2L** algorithm: The First-Fit heuristic in the VM scheduler and MAP-H2L host selection policy to minimize the number of active physical hosts. The MAP-H2L is sorting in descendant of workload from highest to lowest,

(4) **HalfPI-FF-MAP** algorithm: The First-Fit heuristic in the VM scheduler and MAP host selection policy to minimize the number of active physical hosts, and combined with the consolidation of the VM on the ratio θ of consolidation is π/2 (≈1.57),

(5) **PI-FF-MAP** algorithm: similar to HalfPI_FF_MAP, except that in this case we set the ratio θ of consolidation is π (≈3.14).

The result is summary in Table 1 and Fig. 2. We assume that power calculate as in (4), such that a simulated physical hosts with full CPU utilization has power is 250W ($P_{max}$=250W, k=0.7).

*Table 1*. Experiment study on four VM allocation algorithms

| VM Scheduling/Allocation algorithm | Energy (KWh) | QoS | Ratio θ | Decrease in Energy (KWh) | (%) |
|---|---|---|---|---|---|
| (1) OriginalGreedy | 1610.51 | 100% | 1.00 | 0 | 0% |
| (2) FF-MAP | 1493.88 | 100% | 1.00 | 116.63 | 7.24% |
| (3) FF-MAP-H2L | 1490.97 | 100% | 1.00 | 119.54 | 7.42% |
| (4) HalfPI-FF-MAP | 1049.00 |  | 1.57 | 561.51 | 34.87% |
| (5) PI-FF-MAP | 593.90 |  | 3.14 | 1016.61 | 63.12% |

When QoS is still at 100% (i.e. ratio θ of consolidation is 1.00), there are OriginalGreedy, FF-MAP-H2L and FF-MAP algorithms. The FF-MAP and FF-MAP-H2L algorithms give better than Haizea's original greedy algorithm in reducing energy consumption. Even though if the VM scheduler allows the ratio θ, more VMs can put into a physical host and so the total energy consumption is decreased significantly (up to 34% and 63% in HalfPI-FF-MAP and PI-FF-MAP). The ratio θ of consolidation can apply for cases in low QoS requests.

Fig. 2 shows that our four proposed algorithms (FF-MAP, FF-MAP-H2L, HalfPI-FF-MAP, PI-FF-MAP) have lower number of active physical hosts than original greedy algorithm in [1]. The cloud system can work only the active physical hosts and other passive physical hosts can be on stand-by mode or power off.

We have also validated our two First-Fit algorithms (FF-MAP and FF-MAP-H2L) with increasing cores from 8 to 16 and 32 cores in each physical host. We assume that $P_{max}$ and k=0.7 are unchanged. The results in experiment have showed nearly linear decrease in total energy consumption (in percent) of two First-Fit heuristics with increasing by number of cores in CPU (See the Table 2).

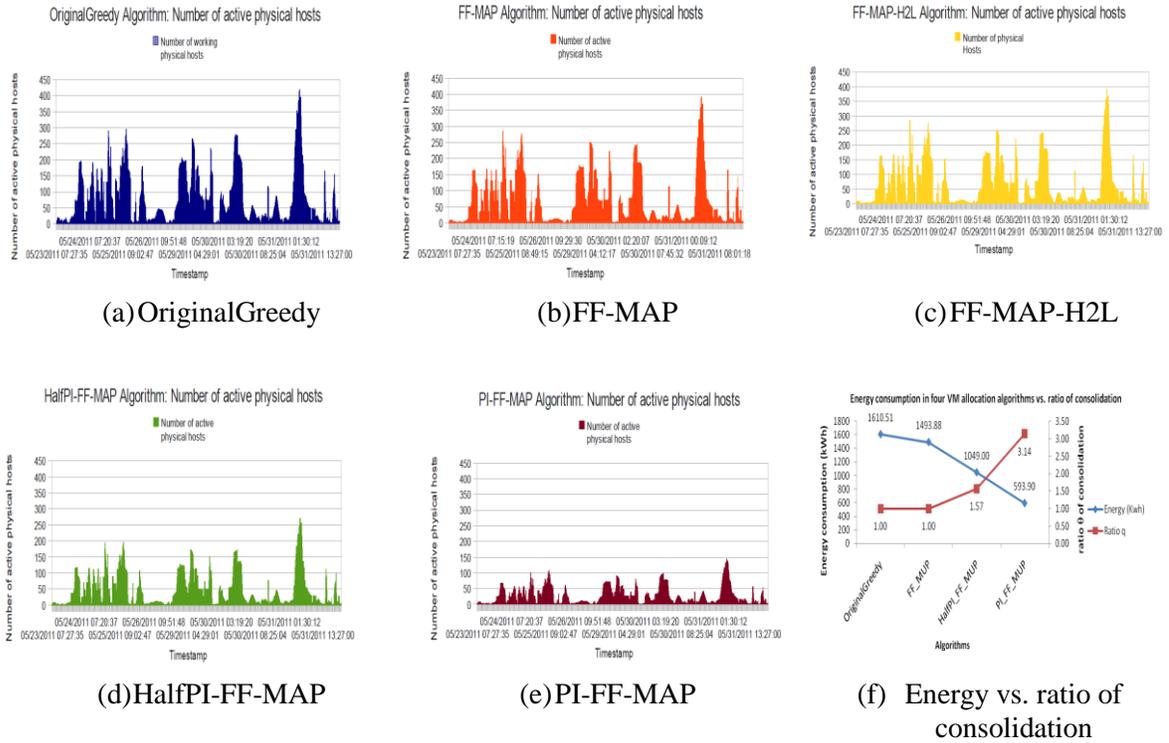

(a) OriginalGreedy  (b) FF-MAP  (c) FF-MAP-H2L

(d) HalfPI-FF-MAP  (e) PI-FF-MAP  (f) Energy vs. ratio of consolidation

*Figure 2*. Number of active physical hosts on the four algorithms (a-e); energy consumption in four VM allocation algorithms vs. the ratio of consolidation (f).

*Table 2*. Evaluating the FF-MAP and FF-MAP-H2L algorithms in multi-core host

| Algorithm | Percent (%) of decrease in total energy consumption | | |
| --- | --- | --- | --- |
|  | 8-core CPU (*) | 16-core CPU (**) | 32-core CPU (***) |
| Original Greedy | 0% | 0% | 0% |
| FF-MAP | 7.24% | 14.32% | 29.86% |
| FF-MAP-H2L | 7.42% | 14.46% | 30.71% |

(*) Each physical host has 8-core CPU, 10240 MB of RAM

(**) Each physical host has 16-core CPU, 16384 MB of RAM

(***) Each physical host has 32-core CPU, 40000 MB of RAM

## 8. CONCLUSIONS AND FUTURE WORKS

In this paper, we present a performance constraint and power-aware allocation for virtual machines using two host selection policies (MAP and MAP-H2L) and four First-Fit mapping heuristics (FF-MAP, FF-MAP-H2L, HalfPI-FF-MAP, PI-FF-MAP). The simulation showed that our four power-aware algorithms could decrease energy consumption on the active physical hosts, and the both FF-MAP and FF-MAP-H2L algorithms compares equally to Haizea's original greedy allocation on two criteria: waiting time and slowdown. The FF-MAP and FF-MAP-H2L algorithms decreased the energy consumption than the original greedy mapping algorithm in Haizea. Moreover, the cloud system can save more energy (KWh) if the VM scheduler accepts on the ratio θ of consolidation. The simulation showed system could decrease in energy consumption significantly. The effect of the ratio of consolidation will make falling in quality to user lease requests.

The determination of power consumption for a data-center accuracy requires that the amount of power consumed by devices such as the air conditioning, the UPS, the average consumption in transition from AC-DC power, network devices (Hub, Switch, Router, ...), etc.

In the future works, we will investigate on workload that combines both best-effort and advanced reservation leases. We also concern on deadline metrics for tasks running on the virtual machines; consider on impact of memory in energy model; estimate performance on the competition of different kind of applications are running on virtual machines which is assigned in same physical computer.